\documentstyle[prb,aps,psfig,twocolumn]{revtex}
\begin{document}
\topmargin -15mm

\title{Implementation of the LDA+U method using the full potential
linearized augmented plane wave basis }

\author{A. B. Shick$^a$, A. I. Liechtenstein$^b$ and W.E. Pickett$^a$ \\
$^a$ Department of Physics,
University of California, Davis, CA 95616 \\
$^b$ IFF, Forschungszentrum J\"ulich, Germany}

\maketitle

\abstract{We provide a straightforward and efficient procedure to
combine LDA+U total energy functional
with the full potential linearized augmented plane wave method.
A detailed derivation of the LDA+U Kohn-Sham type equations
is presented
for the augmented plane wave basis set, and a simple ``second-variation''
based procedure for self-consistent LDA+U calculations is given.
The method is applied to calculate electronic structure and magnetic
properties of NiO and Gd. The magnetic moments and band eigenvalues obtained are in very
good quantitative agreement with previous full potential LMTO calculations.
We point out that LDA+U reduces the total d charge on Ni by 0.1 in NiO.}


\section{Introduction}

The limitations of the local density approximation (LDA) in
calculations of electronic and magnetic ground state properties
of strongly correlated materials are well known \cite{Anisimov}.
One of the simplest methods for going beyond
LDA is provided by so-called ``LDA+U'' total energy functional
\cite{Anisimov,Licht}, which has proven to give large but reasonable
corrections for a number of magnetic insulators. 
(Throughout this paper we understand LDA to include its generalization
to spin polarized systems.)
The method presumes that an appropriate set of local orbitals $\{ \phi_m \}$ (such as 
$3d$ or $4f$) can be identified.  Then a strong intra-atomic
interaction is introduced and treated in a Hartree-Fock-like manner;
in particular, the intra-atomic exchange term is treated without any
local density approximation.  A primary result is the splitting apart
of occupied and unoccupied states within the $\{ \phi_m \}$ shell.

Technically, the explicit use of local orbitals and an 
orbitally dependent correction to the energy functional and 
to the LDA effective potential makes it most convenient to implement
the LDA+U within a method using atomic-like orbitals as basis functions.
This is true for the linearized muffin-tin orbital method (LMTO)
in atomic sphere approximation (ASA) \cite{Anisimov}
or with its full potential
version \cite{Licht}, both of which are formulated in terms of atom-based
orbitals, and most LDA+U work has been done with the LMTO-ASA method.  
Recently the implementation of LDA+U
in the pseudopotential plane wave method was presented by Sawada\cite{Sawada}
by projecting the plane waves onto an atomic orbital.

The LDA+U method has not yet been implemented within an all-electron,
full potential method whose basis set is not based explicitly on
local orbitals.   The linearized augmented plane wave (LAPW) method,
which makes no shape approximation and is acknowledged to be
state-of-the-art in accuracy, uses a basis set of plane waves that
are matched  onto a linear combination of all radial solutions 
(and their energy derivative) inside a sphere centered on each atom.
Hence the basis functions are nothing like individual atomic orbitals.
It is important to establish that the LDA+U method is not limited to
certain basis sets, but rather is more widely applicable.
The full potential aspect may also become important in delicate
situations, such as at the onset of orbital ordering \cite{Licht}.
In this paper we present a simple and numerically efficient
procedure to combine the LDA+U total energy functional with the full potential
linearized augmented plane wave (FLAPW) method \cite{Wimmer}. 
As an example of its efficacy
we apply our implementation to study the effects of strong on-site repulsion
on the magnetic
insulator NiO and the ferromagnetic metal Gd and compare with previous
work on these materials.

The paper is organized as follows.
For the sake of completeness, in Sec. II we recall the basic 
equations of LDA+U
method \cite{Anisimov,Licht}. Then we address the fact that the LAPW method is
not a local orbital basis set approach.
We show that we can define the on-site orbital occupation matrix
and formulate a variational procedure to obtain Kohn-Sham single particle
equations. We also describe a ``second variation'' procedure that
may provide additional insight in analyzing the results.
In Sec. III we present the results of self-consistent total
energy LDA+U FLAPW calculations for the antiferromagnetic insulator NiO and
for Gd metal. These results are compared with previous work, and additional
features are pointed out.

\section{Computational method}


The essence of the LDA+U method is to identify atomic-like orbitals $\{ \phi_m \}$
($m$ is the magnetic quantum number, we will often suppress other indices)
and to treat interactions of electrons in
these orbitals in a non-LDA manner.
The LDA+U variational total energy functional takes the form \cite{Licht} :

\begin{equation}
E^{tot}(\rho,\hat{n}) =  E^{LDA}(\rho) + E^{ee}(\hat{n}) - E^{dc}(\hat{n}) \; , 
\label{1}
\end{equation}
where, $E^{LDA}(\rho)$ is usual local spin density functional
of the total electron spin densities
$\rho^{\sigma}({\bf r}) \; ( \sigma = \uparrow, \downarrow )$.
$E^{ee}$ is an electron-electron interaction energy and $E^{dc}$ is
a ``double-counting'' term which accounts approximately
for an electron-electron interaction
energy already included in $E^{LDA}$. Both are functions of
the local orbital occupation matrix $\hat{n}^{\sigma}= n^{\sigma}_{mm'}$
of the orbitals $\{ \phi_m \}$.

A particular form of $E^{ee}$ is taken in accordance to the multi-band
Hubbard model for d(f) electrons \cite{Licht}

\begin{eqnarray}
E^{ee} = \frac{1}{2}  \sum_{m_1,m_2,m_3,m_4}^{\sigma,\sigma'}
n^{\sigma}_{m_1,m_2}
\Big( <m_1,m_3|V^{ee}|m_2,m_4> \\ \nonumber
- <m_1,m_3|V^{ee}|m_4,m_2> \delta_{\sigma,\sigma'} \Big)
n^{\sigma'}_{m_3,m_4}
\label{2}
\end{eqnarray}
where the $V^{ee}$ is an effective on-site ``Coulomb'' interaction and
$<|>$ indicates an angular integral.

The electron-electron interaction potential in the atomic limit
is given by

\begin{eqnarray}
<m_1,m_3|V^{ee}|m_2,m_4> = \sum_{k}&a_k(m_1,m_3,m_2,m_4)&F_k \\ \nonumber
a_k(m_1,m_3,m_2,m_4)    = \frac{4\pi}{2k+1} \sum_{q=-k}^{k}&<lm_1|Y_{kq}|lm_2>& \times
\\ \nonumber
&<lm_3|Y_{kq}^{*}|lm_4>&
\label{3}
\end{eqnarray}
where $F_k$ are the Slater integrals, $|lm>$ is d(f)-spherical harmonic,
and $n^{\sigma}_{m_1,m_2}$ is the on-site d(f) occupation matrix in the spin-orbital space,
which has to be defined with respect to the chosen  localized orbital basis set.
Note that in $E^{ee}$ in Eq.(2) the self-interaction in the first term
($m_1=m_2=m_3=m_4$ and $\sigma = \sigma'$) is cancelled exactly by the exchange interaction
in the second term. This is one of the important features of the LDA+U approach.
The ``double-counting'' term is taken to satisfy an ``atomic-like'' limit
of the LDA total energy \cite{Solovyev}:

\begin{equation}
E^{dc}(\hat{n}) =  \frac{U}{2}n(n-1)
- \frac{J}{2}\sum_{\sigma}n^{\sigma}(n^{\sigma}-1) 
\label{4}
\end{equation}
where the Hubbard $U$ and exchange $J$ constants are given by
\begin{center}
\begin{eqnarray}
U &=& \frac{1}{(2l+1)^2}\sum_{m_1,m_3} <m_1,m_3|V^{ee}|m_1,m_3> \; , \\ \nonumber
J &=& U - \frac{1}{2l(2l+1)}\sum_{m_1,m_3} [<m_1,m_3|V^{ee}|m_1,m_3> \\ \nonumber 
                                             &-& <m_1,m_3|V^{ee}|m_3,m_1>]\; .
\label{5}
\end{eqnarray}
\end{center}

In Eq.(4), $n^{\sigma} = Tr \; \hat{n}^\sigma$, and $n=\sum_{\sigma} n^{\sigma}$
is a total d(f) on-site occupation numbers.
The energy Eq.(1) is a functional with respect to the spin density
\begin{eqnarray}
\rho^{\sigma}({\bf r}) = \sum_{i} f_i \Phi_i^{*,\sigma}({\bf r}) \Phi^{\sigma}_i({\bf r})  
\label{6}
\end{eqnarray}
and the local orbital occupation matrices $\hat{n}^{\sigma}$.

\subsection{Density matrix definition within the LAPW basis}
It is rather straightforward to choose an orbital basis for Eq.(3) in the case
of the LMTO method as an orthogonal atomic-type LMTO basis set \cite{Licht}.
In the case of the LAPW \cite{Wimmer} method
there are no naturally defined atomic-like basis orbitals.
Therefore, we will use a
projection of augmented plane waves onto radial functions inside 
muffin-tin spheres to define the density
matrix for Eq.(3).  

Let us recall some basic features of LAPW method
(we will use the notations of Singh\cite{Singh}).
The LAPW basis function $\phi^{\sigma}_{\bf k+G} ( {\bf r} )$ is a planewave
$\exp \Big( i{\bf (k+G) \cdot r} \Big)$ outside MT sphere, and inside sphere $i\; 
({\bf r}_{i} ={\bf r} 
- {\bf R}_{i})$
it is a linear combination of all radial functions

\begin{equation}
\phi^{\sigma}_{\bf k+G}({\bf r}) \;=\; \sum_{l,m} \;[ a^{lm}_{\bf k+G} u^{\sigma}_l(r_i) +
b_{\bf k+G}^{lm} \dot{u}^{\sigma}_l(r_i) ] Y_{lm}(\hat{r}_i)
\label{7}
\end{equation}
where $u_l,\; \dot{u_l}$ are the radial wavefunction and its energy derivative 
at an appropriate energy $E_{l}$ for angular quantum number $l$
(recall $<u_l|u_l> = 1, <u_l|\dot{u_l}>=0$), $\bf k$ is a k-point in BZ, and
${\bf G}$ is any reciprocal lattice vector. 

The solution of the Kohn-Sham equations is a set of eigenfunctions

\begin{equation}
\Phi^{b,\sigma}_{\bf k}({\bf r}) \;=\; \sum_{\bf G} \; c^{b,\sigma}_{\bf k+G} 
\phi^{\sigma}_{\bf k+G}({\bf r})
\label{8}
\end{equation}
where $b$ stands for band index and the sum runs over reciprocal
lattice vectors.
The charge density is

\begin{equation}
\rho^{\sigma}({\bf r}) \;=\; \sum_{{\bf k}, b} \; f_{{\bf k}b} 
{|\Phi^{*,b,\sigma}_{\bf k}({\bf r})|}^2 
\label{9}
\end{equation}
where $f_{{\bf k},b}$ is the occupation of the state.
In the $i$-th sphere, from Eqs. (8) and (9) the spin density is
\begin{eqnarray}
\rho^{\sigma}({\bf r})\;=\; &\sum_{{\bf k}, b} \; f_{{\bf k}b}
\sum_{\bf G,G'} c^{*,b,\sigma}_{\bf k+G'} c^{b,\sigma}_{\bf k+G}&\times \\ \nonumber
\sum_{l',m',l,m} & [a^{*,l'm'}_{\bf k+G'} u^{\sigma}_{l'}(r_i) +
b_{\bf k+G'}^{*,l'm'} \dot{u}^{\sigma}_{l'}(r_i) ] & \\ \nonumber 
&[a^{lm}_{\bf k+G} u^{\sigma}_l(r_i)+
b_{\bf k+G}^{lm} \dot{u}^{\sigma}_l(r_i) ] &
Y_{lm}(\hat{r}_i) Y^{*}_{l'm'}(\hat{r}_i).
\label{10}
\end{eqnarray}
Taking the elements of the density matrix from the projection of the
wavefunctions onto the $Y_{lm}$ subspace gives 

\begin{eqnarray}
n^{\sigma}_{lm,l'm'} \;=\; && 
\sum_{{\bf k}, b} \; f_{{\bf k}b}
\sum_{\bf G,G'} \; c^{*,b,\sigma}_{\bf k+G'} c^{b,\sigma}_{\bf k+G} \times \\ \nonumber
\int drr^2&&\;[ a^{*,l'm'}_{\bf k+G'} u^{\sigma}_{l'}(r_i) +
b_{\bf k+G'}^{*,l'm'} \dot{u}^{\sigma}_{l'}(r_i) ] \times \\ \nonumber 
&&[ a^{lm}_{\bf k+G} u^{\sigma}_l(r_i) +
b_{\bf k+G}^{lm} \dot{u}^{\sigma}_l(r_i) ]
\label{11}
\end{eqnarray}
Such a projection gives the desired result (i.e. in agreement with other
implementations) in the atomic limit, and its diagonal elements give what
are commonly thought of as local orbital occupation factors.

This expression indicates that the density matrix is not necessarily
diagonal in $l$ but this detail is not expected to be important.
Keeping only the $l'=l$ part and suppressing the
$l$ index on the density matrix, after radial integration Eq.(11) becomes

\begin{eqnarray}
n^{\sigma}_{m,m'} \;=\; 
\sum_{{\bf k}, b} \; f_{{\bf k}b}
\sum_{\bf G,G'} \; c^{*,b,\sigma}_{\bf k+G'} c^{b,\sigma}_{\bf k+G} \times \\ \nonumber 
[a^{*,lm'}_{\bf k+G'} a^{lm}_{\bf k+G} + 
b_{\bf k+G'}^{*,lm'}  
b_{\bf k+G}^{lm} <\dot{u}|\dot{u}>]
\label{12}
\end{eqnarray}
Finally, one can rewrite Eq. (12) as

\begin{eqnarray}
n^{\sigma}_{m,m'} \;=\;
&& \sum_{{\bf k}, b} \; f_{{\bf k}b} [
<u^{\sigma}_{l} Y_{lm'}|\Phi_{\bf k}^{b,s}><\Phi_{\bf k}^{b,s}|u^{\sigma}_{l} Y_{lm}> + \\ \nonumber
&&<\dot{u}^{\sigma}_{l} Y_{lm'}|\Phi_{\bf k}^{b,\sigma}><\Phi_{\bf k}^{b,\sigma}
|\dot{u}^{\sigma}_{l} Y_{lm}>/<\dot{u}|\dot{u}>]
\label{13}
\end{eqnarray}

\subsection{Variational principle.}

Using Eq.(13) and
orthonormality of the $\Phi$'s one can 
minimize Eq.(1) with respect to $\Phi_{i}^{*,\sigma}$ \cite{Zunger}:

\begin{eqnarray}
\frac{\delta E}{\delta \Phi_{i}^{*,\sigma}}
- \frac{\delta \sum e_j f_j <\Phi^{\sigma}_{j}|\Phi^{\sigma}_{j}>}{\delta{\Phi^{*,\sigma}_i}} = 0
\label{14}
\end{eqnarray}
It gives the set of Kohn-Sham equations,
which can be written in the form

\begin{eqnarray}
\Big( -\nabla^{2} + V^{\sigma}_{LDA}({\bf r}) \Big) \Phi^{\sigma}_i({\bf r}) + \\ \nonumber 
\sum_{m,m'} V^{\sigma}_{mm'} \frac{\delta n^{\sigma}_{m,m'}}{\delta \Phi_i^{*,\sigma}} 
= e^s_{i} \Phi^{\sigma}_{i}({\bf r})
\label{15}
\end{eqnarray}
where an effective potential acting on the $Y_{lm}$ subspace
(the d(f)-states) is 

\begin{eqnarray}
V^{\sigma}_{mm'} = \sum_{p,q,\sigma'}
\Big( <m,p|V^{ee}|m',q> - \\ \nonumber
<m,p|V^{ee}|q,m'> \delta_{\sigma,\sigma '} \Big)
n^{\sigma'}_{p,q} - \\ \nonumber
\delta_{m,m'} U (n-\frac{1}{2})
+ \delta_{m,m'} J (n^{\sigma}-\frac{1}{2})
\label{16}
\end{eqnarray}
and from Eq.(13)

\begin{eqnarray}
&\frac{\delta n^{\sigma}_{m,m'}}{\delta \Phi_i^{*,\sigma}} = 
<u^{\sigma}_{l'} Y_{lm'}|\Phi_{i}^{\sigma}> u_{ls}(r) Y_{lm}(r) + & \\ \nonumber 
&<\dot{u}^{\sigma}_{l} Y_{lm'}|\Phi_{i}^{\sigma}> \dot{u}^{\sigma}_{l}(r)
Y_{lm}(r)/<\dot{u}|\dot{u}> = & \\ \nonumber
& \Big[ |u^{\sigma}_l Y_{lm} > < u^{\sigma}_l Y_{lm'}| +
\frac{1}{<\dot{u}|\dot{u}>}|\dot{u}^{\sigma}_l Y_{lm} >
<\dot{u}^{\sigma}_l Y_{lm'}| \Big] \Phi^{\sigma}_i &
\label{17}
\end{eqnarray}
Eqns (15)-(17) indicate specifically how the additional potential that arises
in the LDA+U method acts on the projections of the wavefunctions $\Phi$
onto the $Y_{lm}$ subspace.

\subsection{Use of ``second variation'' procedure.}

The Eq.(15) can be solved directly using the LAPW basis set Eq.(7).
For large systems it can be solved more efficiently using 
a ``second variation'' procedure.
We define an auxiliary orthogonal basis set 
$\Psi^{b,\sigma}_{\bf k}({\bf r})$ as the solutions of 
LDA band Hamiltonian

\begin{eqnarray}
(-\nabla^{2} + V^{\sigma}_{LDA}({\bf r}))\Psi^{b,\sigma}_{\bf k}({\bf r}) 
= \epsilon^{b,\sigma} \Psi^{b,\sigma}_{\bf k}({\bf r})
\label{18}
\end{eqnarray}
The functions $\Psi_{\bf k}$ are somewhat like the LDA eigenfunctions,
but they are not identical because it is not the LDA density that goes into
$V^{\sigma}_{LDA}$.
Expanding the functions $\Phi$ via $\Psi$ (suppressing {\bf k} and shortening notation)

\begin{eqnarray}
|\Phi^i> = \sum_{j} d^{i}_j |\Psi^j>
\label{19}
\end{eqnarray}
Eq.(15) is then transformed to the following,

\begin{eqnarray}
&& \sum_j \epsilon_{j} d^i_j |\Psi^{j}> 
+ \sum_j d^i_j \sum_{m,m'} V^{\sigma}_{mm'} [<u^{\sigma}_{l} Y_{lm'}|\Psi^{j}> \times \\ \nonumber
&& u^{\sigma}_{l}(r) Y_{lm}(r) +
<\dot{u}^{\sigma}_{l} Y_{lm'}|\Psi^{j}> \dot{u}^{\sigma}_{l}(r) Y_{lm}(r)/<\dot{u}|\dot{u}>] \\ \nonumber
&& = e_{i} \sum_j d^i_j |\Psi^{j} > 
\label{20}
\end{eqnarray}
and using $<\Psi_i|\Psi_j> = \delta_{ij}$ we obtain the ``second variation'' 
Hamiltonian
($\sum_{j'} H_{jj'} d^{i}_{j'} = e_{i} d^{i}_j$) of the form:

\begin{eqnarray}
H^{\sigma}_{j'j} \; = \; \epsilon_{j} \delta_{j'j} \;+\; && \\ \nonumber
\sum_{m,m'} [ <\Psi^{j'}| u^{\sigma}_{l} Y_{lm}>
&V^{\sigma}_{m,m'}&
<u^{\sigma}_{l} Y_{lm'}|\Psi^{j}> + \\ \nonumber
<\Psi^{j'}|\dot{u}^{\sigma}_{l} Y_{lm}>
&V^{\sigma}_{m,m'}&
<\dot{u}^{\sigma}_{l} Y_{l'm'}|\Psi^{j}>/<\dot{u}|\dot{u}>]
\label{21}
\end{eqnarray}

Now, given some plane wave cutoff as usual,
the self-consistent solution of Eq.(15) is performed as follows.
(i) Solving Eq.(18) \cite{Singh} for the given LDA effective 
potential we obtain
the entire orthogonal basis set \{$\Phi$\}; 
(ii) within this basis set we solve
Eq.(21) to obtain the coefficients $d^i_j$ of the LDA+U wavefunctions
in terms of the auxiliary wavefunctions in Eq.(19); 
(iii) the new LDA+U wavefunction is then 
projected back to LAPW basis set (cf., Eq.(7)) and the usual procedure
to calculate charge and spin densities Eq.(10) \cite{Singh} and occupation
matrices Eq.(12) is then used to achieve self-consistent solution.

\subsection{Total energy evaluation}
After the self-consistent solution of Eq. (15)
the LDA+U total energy can be calculated from Eq. (1),
using:

\begin{equation}
E^{LDA}(\rho) =  T_s(\rho) \; + \; \int d {\bf r} v_{ext}({\bf r}) \rho ({\bf r})
\; + \; E_H (\rho) \; + \; E_{xc} (\rho)
\label{22}
\end{equation}
where $T_s$ is the kinetic energy for non-interacting electrons
with density $\rho$,
$v_{ext}$ is an external potential (including the interaction with nuclei), 
$E_H$ is the Hartree energy and $E_{xc}$ is an exchange-correlation
energy. It is convenient to use, as usual \cite{Singh},
the eigenvalue sum, which includes $T_s$ and the interaction with nuclei
exactly but miscounts other terms.
A so-called ``double-counting'' correction $E^{dc}_{LDA}$ leads to the correct
energy.

For our LDA+U calculations, the eigenvalue sum contains additional terms
(involving U and J) whose contributions to the energy are miscounted.
When these terms are corrected, the expression for the energy is:
\begin{equation}
E^{tot}(\rho,\hat{n}) =  \sum^{occ}_{i} \epsilon_i - E^{dc}_{LDA}(\rho)
- E^{dc}_{LDA+U}(\hat{n})
\label{23}
\end{equation}
where,
$E^{dc}_{LDA}$ is a ``double-counting'' LDA correction,
and $E^{dc}_{LDA+U}$ is a ``double-counting'' correction to the
eigenvalue sum in LDA+U calculations. This correction is expressed
in terms of electron-electron interaction energy Eq. (2) and 
``double-counting'' term Eq.(4) as
\begin{equation}
E^{dc}_{LDA+U}(\rho,\hat{n}) =   E^{ee}(\hat{n}) - E^{dc} (\hat{n})
- \frac{1}{2}(U-J)n 
\label{24}
\end{equation}
where, the last term appears due to a specific form of the LDA+U
``double counting'' energy Eq. (4) with an explicit removal
of self-interaction.
 
The LDA+U expression for the energy (Eq. (1)) has been used very sparingly
except to generate the Kohn-Sham equations which lead to the self-consistent
density and orbital occupation matrices. Partly this has been due to the 
implementation of the LDA+U in the LMTO-ASA method, which makes severe 
approximations to the shape of the potential and density and for the
variational freedom of the wavefunctions, which translates into questionable
ability to resolve energy differences. Another feature of the LDA+U energy
functional that has not been emphasized is that it explores a regime of 
$E_{LDA}[\rho]$ which is not understood, by evaluating $E_{LDA}$ for densities
that do not minimize it. This procedure deserves scrutiny, so we will give attention
to the energy differences in the following section.

\section{Results}

As representative systems to illustrate the numerical procedure we
have proposed, we choose
the  antiferromagnetic insulator NiO and Gd metal,
which has a ferromagnetic ground state.
For both of them LDA fails to provide
the correct electronic and magnetic ground states due to significant
electron correlation effects.

\subsection{NiO}
LDA-ASA\cite{Terakura,Anisimov} calculations of
antiferromagnetic (AFII) NiO gives the value of the band gap
and magnetic moment in disagreement
with experimental data (cf., Table \ref{tab1}).
While the band gap requires an energy dependent self-energy
to obtain rigorously and therefore is not a true test of LDA+U
corrections, the magnetic moment is a fundamental test. Generally,
it is expected that an improved energy functional will also give a band gap
in closer agreement with experiment than the very poor LDA value.
Our FLAPW LDA calculations give somewhat larger values of both
moment and gap than does the ASA calculation (Table \ref{tab1}),
reflecting the approximations in the shape of the density and potential,
and the less flexible basis functions in the ASA. 
For NiO, we choose the lattice constant
as 7.927 a.u. in accordance to the experimental data \cite{Dudarev}.
For self-consistency, 110 special k-points\cite{BZ} in the 
irreducible part of the AFII BZ were used,
with Gaussian smearing of 2 mRy of the Fermi energy to promote convergence.
The sphere radii $R^{Ni}_{MT} = 2.0 \; a.u. $ 
 and $R^{O}_{MT} = 1.8 \; a.u.$ were used, and
$R^{O}_{MT} \times K_{max} = 6.6 $ determined the basis set size.
Literature values $U \; = \; 8.0 $ eV and $J \; = 0.95$ 
eV\cite{Dudarev} were used to obtain
the values of Slater integrals in Eq.(3)
($F_0 =$ 8.00 eV, $F_2 =$ 8.19 eV, $F_4 =$ 5.11 eV).

The calculated values of spin magnetic moment and band gap are shown in Table
\ref{tab1},
are within about 8\% of recent
full-potential LMTO based LDA+U calculations \cite{Dudarev}. 
The calculated projected densities of states (PDOS) for Ni d states and
O p states are shown in Fig. 1
and are very similar to these of Ref. \cite{Dudarev}. The $t_{2g}$ and $e_{g}$
PDOS for Ni d states for LDA+U are compared with those for LDA in Fig. 2.
The LDA+U calculations give a charge transfer type
band gap ($\approx$ 3.4 eV) between mainly O p states at the top
of valence band and primarily Ni $e_g$ states in the conduction band as opposed to
the LDA calculations where a small band gap (0.4 eV) is formed between mainly
Ni d states. There is a large downward shift of majority Ni $e_g$ states (cf. Fig. 2)
and the resulting energy difference between $e_{g}^{\uparrow}$ and $e_{g}^{\downarrow}$
states ($\approx$ 10.8 eV) is not far from model GW calculations 
($\approx$ 9 eV) \cite{Massida}.
For the symmetry of the Ni ion, the occupation matrix is specified
by two numbers (for each spin), the $e_g$ and $t_{2g}$ occupancies.
The LDA+U correction results in a decrease in the Ni d occupation by 0.12
electrons, with an increase in spin majority charge of 0.2 electrons
(to essentially 100\%) more
than compensated by a decrease by 0.3 electrons in spin minority charge
(nearly all of it $e_g$). This difference of d charge is potentially
important for the understanding of correlation in magnetic insulators.  

Since we take into account only the muffin-tin part of LDA+U correction,
it is important to analyze numerically the stability of our results with
respect to the choice of the MT radii. We have varied the sphere radii for
both Ni and O atoms in the unit cell and found that these changes have 
little effect on
the calculated values of magnetic moment and band gap
(cf. Table {\ref{tab2})), due to the fact that the sphere contains all but
a few percent of the Ni d states.

\subsection{Gd}
A detailed analysis of ground state and magnetic properties of ferromagnetic
Gd has been presented by Singh.\cite{Singh} It was shown 
that Gd 4f-electrons have
to be considered as ``band'' electrons instead of ``core'' to obtain at least
qualitative agreement with experimentally observed Fermi-surface. 
(The ``core'' treatment consists of treating the 4f states as corelike,
nondispersive bands that do not hybridize with the valence bands, and 
removing the 4f character from the valence electron basis set.)  Heinemann and Temmerman
found with the LMTO-ASA method \cite{temmerman} that using GGA rather than
LDA led to a ferromagnetic state being favored over the AF state. The more
recent full potential LMTO calculations of Harmon {\it et al.}\cite{Harmon} did 
not confirm this conclusion.
Instead, it was shown that only LDA+U gives a correct ferromagnetic ground
state for Gd.

For Gd, we chose the lattice constant
as 6.8662 a.u. and c/a ratio as 1.587
in accordance with the
experimental crystal structure data \cite{struc}.
Here,  150 special k-points in irreducible part of the hexagonal BZ were used,
again with Gaussian smearing.
The values $R_{MT} = 3.2 \; a.u. $ and
$R_{MT} \times K_{max} = 9.6 $ were used.
Literature values of Ref. \cite{Harmon} $U \; = 6.7 $ eV and $J \; = 0.7 $eV 
were used to obtain
the values of Slater integrals in Eq.(3)
($F_0 =$ 6.70 eV, $F_2 =$ 8.34 eV, $F_4 =$ 5.57 eV, $F_6 =$ 4.13 eV ).

The spin magnetic moment for FM and AFM Gd are shown in Table \ref{tab3}
for both LDA and LDA+U calculations. The LDA result was in nearly perfect
agreement with experiment, and the 2\% increase given by LDA+U does not
degrade the agreement much.
The energy difference $E(AFM) - E(FM)$ is negative in LDA calculations and positive
in LDA+U calculations.
Our results therefore confirm quantitatively the conclusion of Ref. \cite{Harmon}
that the correct FM result is only predicted by LDA+U. 
The LDA and LDA+U partial f-densities of states are shown in Fig. 3.
There is a large (4.5 eV) shift of spin-majority f-states below the Fermi level
and removal of the unoccupied spin-minority f-states from close vicinity of the Fermi
level due to an upward shift of 1.5 eV.
The enhancement of an exchange splitting can be understood as an increase of
an effective exchange splitting parameter $I \; = \; (U \; + \; 6J)/7 \; = \; 1.5 \; eV$
in mean-field Hubbard model in comparison with LDA ($I \; = \; J \; = \; 0.7 \; eV$)
due to Hubbard $U$.
The exchange splitting ($IM$) for f-states is then increases from $5 \; eV$ with LDA to
$11 \; eV$ with LDA+U (cf. Fig.3).
A more complete discussion of the
LDA+U description of Gd will be given elsewhere.

\section{Conclusion}
To summarize, we have presented a straightforward and 
numerically efficient procedure
to combine the LDA+U total energy functional with the FLAPW method. This implementation
is all-electron and includes no shape approximations.  In spite
of a modest limitation -- we use the only ``muffin-tin'' part of the LDA+U
correction -- the method works well for representative systems NiO and
Gd for the electronic spectrum and predicts the correct magnetic order.
It also can be easily extended to incorporate spin-orbit interaction
\cite{Shick}. Moreover, a similar numerical procedure can be used
to implement the ``dynamic
mean field'' parametrization within the FLAPW method as a more 
realistic model for 
the self-energy\cite{Katsnelson} of metallic strongly correlated systems.

We have emphasized the eigenvalue shifts in NiO and in Gd that result from the strong
repulsion in the LDA+U method. These shifts are probably important in obtaining
an improved description. Hovewer, it should be kept in mind that the LDA+U method
is based on the energy functional for the electronic ground state, and that
eigenvalues are not the most fundamental quantaties for judging the success
of the LDA+U method. It is satisfying that the LDA+U energies for Gd lead to the
observed magnetic ground state.

We are grateful to I. V. Solovyev and A. J. Freeman
for helpful comments, and R. Weht and C. O. Rodriguez for
discussion.
This research was supported by
National Science Foundation Grant DMR-9802076.

\newpage

\begin{table}
\caption
{The spin magnetic moments ($M_s$ in $\mu_B$) and band gap ($E_g$ in $eV$) for NiO.
The increases in both quantaties are about the same in the FLAPW and LMTO-ASA methods,
althrough the individual value differ due to approximations made in the LMTO-ASA method.}
\begin{tabular}{cccccccc}
           &\multicolumn{2}{c}{FLAPW}&\multicolumn{2}{c}{LMTO-ASA \cite{Anisimov}}& FLMTO \cite{Dudarev}& Exp. \cite{Anisimov} \\
           &  LDA & LDA+U & LDA & LDA+U & LDA+U  &           \\
$M_s$      &1.186 & 1.687 & 1.0 & 1.56  & 1.74   & 1.64-1.70 \\
$E_g$      &0.41  & 3.38  & 0.2 & 3.1   & 3.4    & 4.0 - 4.3 \\
\end{tabular}
\label{tab1}
\end{table}

\begin{table}
\caption
{The spin magnetic moments ($M_s$ in $\mu_B$) and band gap ($E_g$ in $eV$)
for different ``muffin-tin'' radii (a.u.)  for Ni and O atoms in NiO.
The small differences are entirely unimportant.}
\begin{tabular}{cccccccc}
Ni         & 2.0  & 2.1  & 2.1  \\
O          & 1.8  & 1.8  & 1.7  \\
\hline
$M_s$      & 1.687& 1.700&1.700 \\
$E_g$      & 3.38 & 3.38 & 3.38 \\
\end{tabular}
\label{tab2}
\end{table}

\begin{table}
\caption
{The spin moments (in $\mu_B$) and the total energy difference (in $meV$)
between antiferromagnetic (AFM) and ferromagnetic (FM) phases for the bulk Gd.}
\begin{tabular}{cccccc}
           & LDA & LDA+U & Exp. \\
\hline
FM         & 7.659& 7.818 & 7.63 \\
AFM        &7.246 & 7.437 \\
\hline
E(AFM-FM)  &LDA+U & LDA+U \cite{Harmon} & LDA   &LDA-ASA \cite{temmerman} \\
per atom, meV&63& 68   & -2 & $\approx$ -5                  \\
           &      &                     &       & GGA-ASA \cite{temmerman}  \\
           &      &     & & +1.4 \\
\end{tabular}
\label{tab3}
\end{table}

\newpage

\begin{center}
\begin{figure}
\label{fig1}
\psfig{file=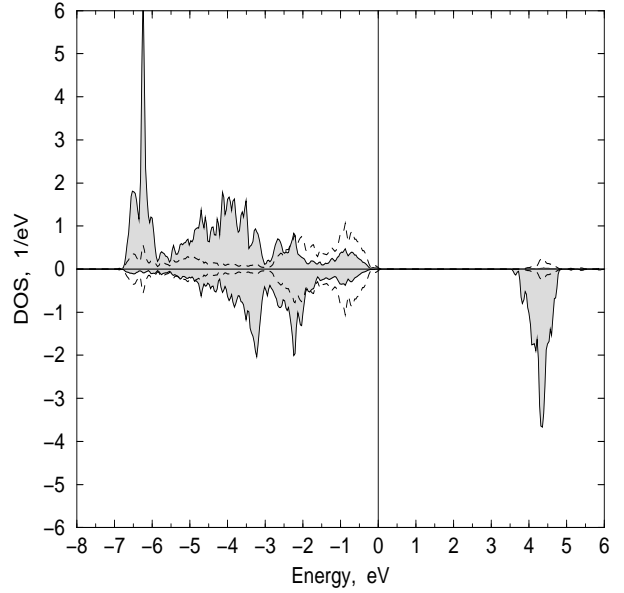,width=8cm,height=8cm}
\caption{ The partial DOS for Ni d-states (filled) and for O p-states for NiO.
Majority is plotted upwards; minority is plotted downward.}
\end{figure}
\end{center}

\newpage

\begin{center}
\begin{figure}
\label{fig2}
\psfig{file=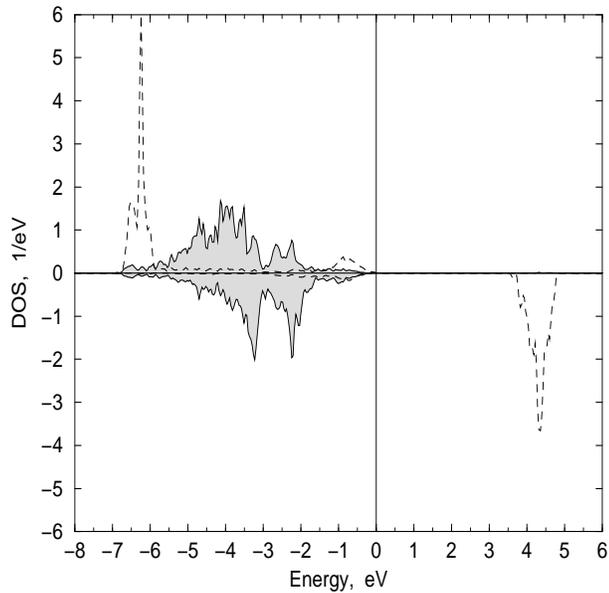,width=8cm,height=8cm}
\psfig{file=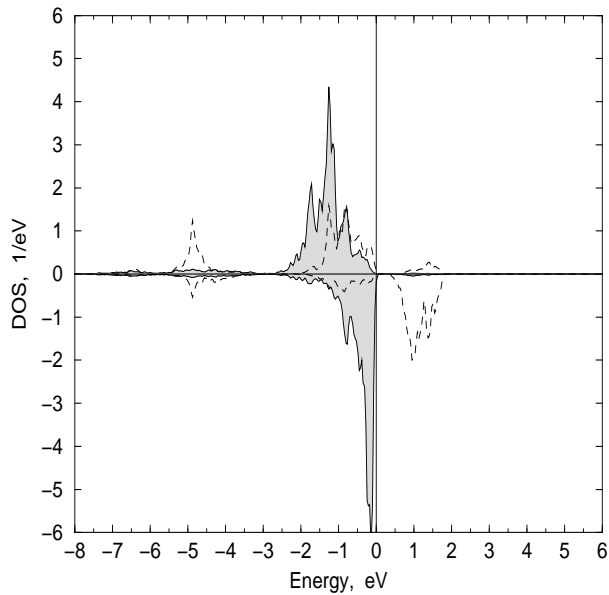,width=8cm,height=8cm}
\caption{ The partial DOS for Ni $t_{2g}$ (filled) and $e_g$ states for LDA+U (a) and
LDA (b) for NiO. The increase in the exchange splitting of the Ni d states due to LDA+U is
dramatic.}
\end{figure}
\end{center}

\newpage

\begin{center}
\begin{figure}
\label{fig3}
\psfig{file=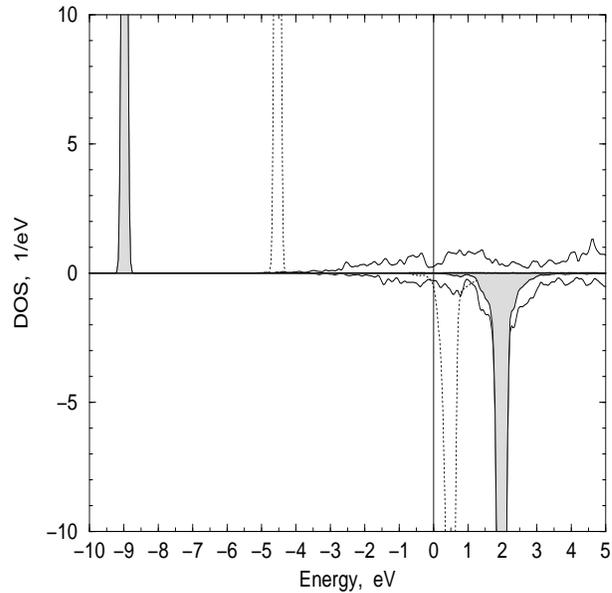,width=8cm,height=8cm}
\caption{ The total DOS (full line), the partial DOS for Gd f-states for LDA+U (filled)
and LDA (dotted) for 
the ferromagnetic Gd. The 4f exchange splitting is increased from
5 eV to 11 eV by LDA+U.
Majority is plotted upwards; minority is plotted downward.}
\end{figure}
\end{center}

\end{document}